\documentclass[referee]{aa}

\usepackage{graphics}

\newcommand{\be}{\begin{equation}}
\newcommand{\ee}{\end{equation}}
\begin{document}

\thesaurus{06    
            (13.07.2; 08.05.01; 09.19.2; 08.16.6)}

\title{Unidentified 3EG gamma-ray sources at low galactic
latitudes}

\author{G.E. Romero \inst{1,}\thanks{Member
of 
CONICET}, P. Benaglia \inst{1,\star}, Diego F. Torres \inst{2}}
\offprints{G.E. Romero}

\institute{Instituto Argentino de Radioastronom\'{\i}a, C.C.5,
(1894) Villa Elisa, Buenos Aires, Argentina \and
Departamento de F\'{\i}sica, UNLP, C.C.67, (1900) La Plata, Buenos Aires,
Argentina}

\date{\today}

\maketitle


\markboth{Romero et al.: Unidentified EGRET sources}{}

\begin{abstract}
We present a study on the possible association of unidentified $\gamma$-ray
sources in the Third EGRET (3EG) catalog with different types of galactic 
objects such as Wolf-Rayet and Of stars, supernova remnants 
(SNRs), and OB associations (considered as pulsar tracers). We have made use 
of numerical simulations of galactic populations of $\gamma$-ray point
sources in order to determine the statistical significance of the positional 
coincidences. New constraints on pure chance association are presented
for SNRs and OB associations, and it is shown that massive stars present 
marginally significant correlation with 3EG sources at a $3\sigma$ level. 

\end{abstract}

\section{Introduction}

The existence of a galactic population of $\gamma$-ray sources 
is known since the days of the COS B experiment (Swanenburg et al. 1981).
Montmerle (1979) showed that about 50 \% of the unidentified COS B 
detections lie in regions containing young objects, like supernova remnants 
(SNRs)
and OB massive stars. He suggested that the $\gamma$-ray emission could 
stem from $\pi^0$-decays resulting from hadronic interactions of high-energy
protons (or nuclei) and ambient matter. These protons would be locally 
injected by young stars in the SNR shocks where they would be diffusively 
accelerated up to high energies by Fermi mechanism. 

Cass\'e \& Paul (1980) argued that particle acceleration at the 
terminal shock of strong stellar winds alone could be responsible
for the $\gamma$-ray sources without the mediation of the SNR shock waves 
advocated by Montmerle. Gamma-ray production in shocks generated by
massive stars has been discussed since then, and from different points of 
view, by V\"olk \& Forman (1982), White (1985), Chen \& White (1991a,b), and 
White \& Chen (1992), among others.

Since 1991, with the advent of the Energetic Gamma Ray Experiment 
Telescope (EGRET) 
onboard the Compton satellite, the observational data on galactic 
$\gamma$-ray sources have been dramatically improved. Two of the previously 
unidentified COS-B sources, Geminga and PSR 1706-44, are now known
to be pulsars. The detection of pulsed high-energy emission from other 
sources (there are seven $\gamma$-ray pulsars so far, see Thompson 1996
for a review) and the identification of Geminga as a radio quiet object 
have prompted several authors to explore the possibility that all 
unidentified low latitude sources in the Second EGRET (2EG) catalog 
(Thompson et al. 1995, 1996) are pulsars (with the
exception of a small extragalactic component which is seen through the 
Galaxy). In particular, Kaaret \& Cottam (1996) have used OB 
associations as pulsar tracers finding out a significant positional
correlation with 2EG unidentified sources. A similar study, 
including SNRs and HII regions (these latter considered as tracers of 
star forming regions and, consequently, of possible pulsar concentrations), 
has been carried out by Yadigaroglu \& Romani (1997), who concluded 
that the pulsar hypothesis for the 2EG sources is consistent with the 
available information.

However, recent spectral analyses made by Merck et al. (1996) and Zhang
\& Cheng (1998) clearly show that several 2EG sources are quite at odds 
with the pulsar explanation. Time variability in the $\gamma$-ray flux of 
many sources also argues against a unique population behind the 
unidentified galactic $\gamma$-ray detections (McLaughlin et al. 1996, 
Mukherjee et al. 1997). 

Sturner \& Dermer (1995) and Sturner et al. (1996) have investigated the 
possible association of $\gamma$-sources with SNRs, finding significant 
statistical support for the idea that some remnants could be
$\gamma$-ray emitters. Esposito 
et al. (1996) have shown that five 2EG sources are coincident with well 
known SNRs and, more recently, Combi et al. (1998a) have detected a new 
shell-type SNR at the position of 2EGS J1703-6302, as well as an interacting
compact HI cloud, through multiple radio observations, clearly demonstrating,
in this way, that at least some EGRET detections are physically related
to SNRs.

With the publication of the Third EGRET (3EG) catalog of high-energy 
gamma-ray sources (Hartman et al. 1999), 
which includes data from Cycles 1 to 4 of the space 
mission, new and valuable elements become available to deepen the quest  
for the nature of the unidentified $\gamma$-ray sources. The new catalog 
lists 271 point sources, including 170 detections with no conclusive 
counterparts at other wavelengths. Of the unidentified sources, 74 are 
located at $|b|<10^o$ (this number can be extended to 81 if we include
sources with their 95 \% confidence contours reaching latitudes $|b|<10^o$).
This means that the number of possible galactic unidentified sources is now
nearly doubled respect to the 2EG catalog.

Can these new sources be associated with pulsars? How many sources could be
ascribed to known SNRs? Is there new statistical evidence for the 
identification of some detections in the 3EG catalog with massive stars 
that generate very strong winds? In the present paper we investigate these 
questions in the light of the new $\gamma$-ray data of the 3EG catalog.     
We use numerical simulations (constrained by adequate boundary conditions)
of $\gamma$-ray source populations to weight the statistical 
significance of the different levels of positional coincidences determined 
for diverse types of candidates such as individual massive stars (Wolf-Rayet 
and Of stars with strong stellar winds), SNRs, and OB associations.

The contents of the paper are as follows. 
In the next section we describe the numerical procedure implemented for
the analyses. Sections 3, 4, and 5 deal with the possible association of
unidentified 3EG sources with stars, SNRs, and star-forming regions 
considered as pulsar tracers, respectively. In Section 6 we present some 
further comments and, finally, in Section 7, we draw our 
conclusions. 

\section{Numerical simulations and statistical results}

With the aim of finding the positional coincidences between 3EG unidentified 
sources at $|b|<10^o$ and different populations of galactic objects, we 
have developed a computer code that determines the angular distance
between two points in the sky, taking into account the positional 
uncertainties in each of them. The code can be used to obtain a list of 
$\gamma$-ray sources with error boxes (here assumed as the 95 \% 
confidence contours given by the 3EG catalog) overlapping  different kinds
of objects, both extended (like SNRs or OB associations) and punctual 
(like stars).

We run the code with the 81 unidentified EGRET sources at galactic 
latitudes $|b|<10^o$ and 
complete lists of Wolf-Rayet (WR) stars, Of stars, SNRs, and OB associations. 
These lists were obtained from van der Hucht et al. (1988), Cruz-Gonz\'alez
et al. (1974), Green (1998), and Mel'nick \& Efremov (1995), respectively.  
We have found that 6 $\gamma$-ray sources of the 3EG catalog are positionally
coincident with WR stars, 4 with Of stars, 22 with SNRs, and 26 with OB 
associations.

\begin{figure}
\resizebox{\hsize}{!}{\includegraphics{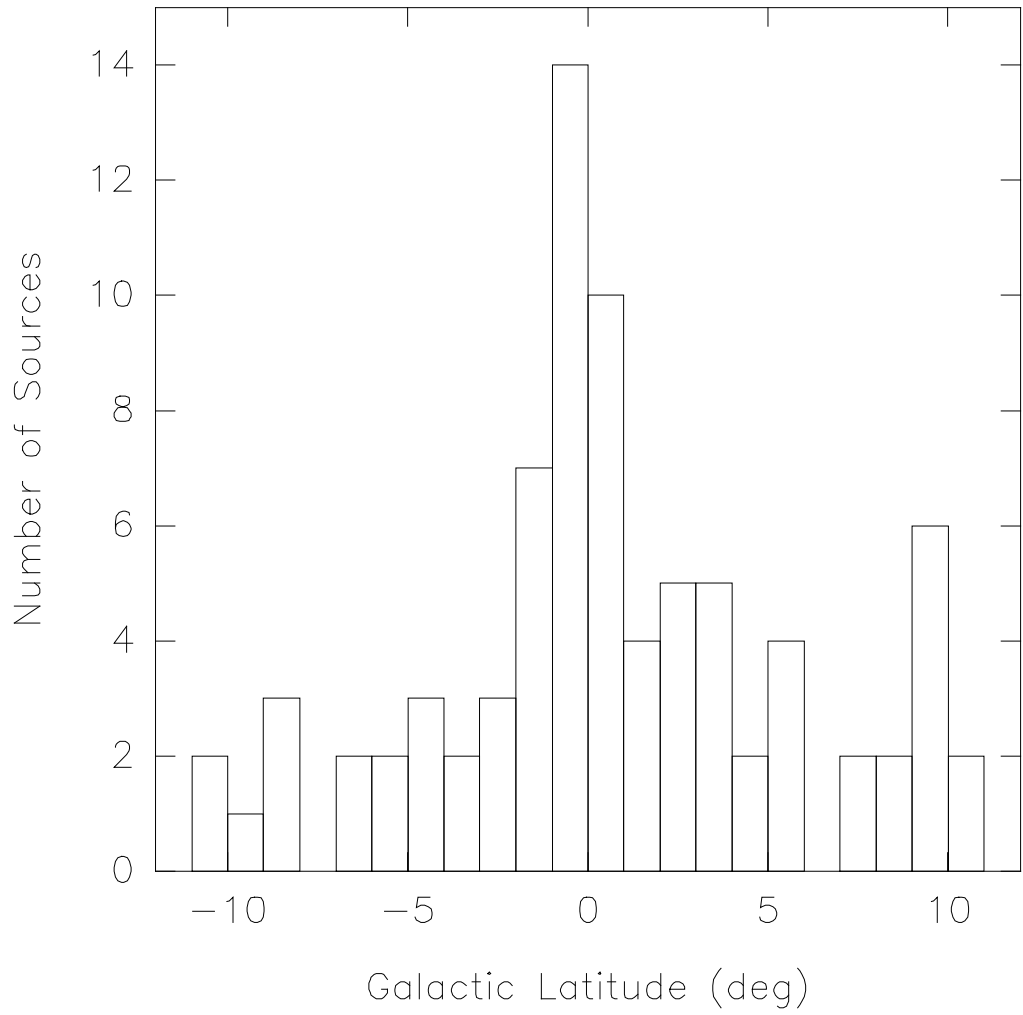}}
\caption{Distribution in glactic latitude of the 81 3EG unidentified EGRET
sources with positions at $|b|<10^o$ (within errors)}
\label{fig.1}
\end{figure}

In order to estimate the statistical significance of these coincidences, 
we have simulated a large number of sets of EGRET detections, retaining for
each simulated position the original uncertainty in its galactic 
coordinates. Specifically, in each case we have generated by computer 1500 
populations of 81 $\gamma$-ray sources through rotations on the celestial 
sphere, displacing a source with original coordinates 
$(l,b)$ to a new position $(l^\prime,b^\prime)$. 
The new pair of coordinates is obtained
from the previous one by setting $l^\prime=l+ R_1 \times 360^o$.
Here, $R_1$ is a random number between 0 and 1, which
never repeats neither from source to source nor from set to set.
Since we are simulating a galactic source population and not arbitrary
sets at $|b|<10^o$, we impose that the new distribution (i.e. each of the 
simulated sets) retains the form of the actual histogram in latitude of 
the unidentified 3EG sources, with 1$^o$ or 2$^o$-binning. The histogram, 
for 1$^o$-binning, is shown in Figure ~\ref{fig.1}.

In order to accomplish the mentioned constraint, we make 
$b^\prime=b + R_2 \times 1^o$, and then, if the integer part of $b^\prime$
is greater than the integer part of $b$ or if the sign of $b^\prime$
and $b$ are different, we replace $b^\prime$ by 
$b^\prime - 1^o$. Here, again, $R_2$ is a random number between
0 and 1. This ensures that the new set of artificial positions preserves
the actual histogram in latitude at 1$^o$-binning. Similarly, a 
2$^o$-binning distribution can be maintained. Both sets of simulations
provide comparable results.

The unidentified 3EG sources have, additionally, a non-uniform
distribution in galactic longitude, showing a concentration towards
the galactic center. However, when doing the simulations, we imposed no
constraints in longitude because we wanted to consider any kind of 
possible galactic populations.

Once we performed 1500 simulations for each type of counterparts (a larger
number of simulations do not significantly modify the results), we 
estimated the level of positional coincidences between each simulated 
set and the different galactic populations under consideration. From these 
results we obtained an average expected value of chance associations and a 
corresponding standard deviation. The 
probability that the observed association level had   happened by chance 
was then evaluated assuming a 
Gaussian distribution of the outputs. The results of this study are 
shown in Table 1, where we list, from left to right, the type of 
object under study, the number of actual positional coincidences,
the number of expected chance coincidences according to 1$^o$-binning 
simulations, the probabilities that the actual coincidences can be due 
to chance, and the similar results for simulations with 2$^o$-binning.

From Table 1, it can be seen that there is a strong statistical correlation 
between unidentified $\gamma$-ray sources of the 3EG catalog and SNRs 
(at $\sim6\sigma$ level) as well as with OB associations (at  
$\sim4\sigma$ level). Regarding the stars, we find that there is a 
marginally significant correlation with WR and Of stars 
($\sim3\sigma$). Remarkably, the probability of a pure chance association 
for SNRs is as low as 5.4$\times10^{-10}$ according to the 2$^o$-binning
simulations ($1.6\times10^{-8}$ for 1$^o$-binning). For the stars, we obtain probabilities in the range 
$10^{-2}-10^{-3}$, which are suggestive but not overwhelming. 

In the next sections we explore these results in more detail.

\section{Massive stars}

The case for possible association of unidentified EGRET
sources with WR stars was previously presented --using data from
the 2EG catalog-- by Raul \& Mitra (1997). In the former catalog, there are 
37 unidentified sources at $|b|<10^o$.
Raul and Mitra proposed, on the basis of
positional correlation, that 8 of these sources could be produced
by WR stars. Their analysis of the possible chance occurrence of these
associations, which was purely analytic and assumed equiprobability for
each position on the sky, yielded an a priori expectation of $\sim 10^{-4}$.
Their results are notably modified when the 3EG catalog is considered.
Changes in position and smaller positional uncertainties reduce the 
number of positional coincidences despite the remarkable increment in 
the number of sources. Additionally, a more rigorous treatment in the 
probability analysis has the effect of significantly enhance the 
possibility of chance association (see Table 1).

In Tables 2 and 3 we list the 3EG sources positionally coincident with
WR and Of stars, respectively. As far as we are aware this is the first 
time that a statistical study of the correlation between Of stars and EGRET 
detections is carried out, despite that the possibility of $\gamma$-ray 
production in this kind of objects has been extensively discussed in the 
literature (e.g. V\"olk \& Forman 1982). In the tables we provide, from 
left to right, the 3EG source name, the measured (summed over Cycles 1 to 4)
$\gamma$-ray  flux, the photon spectral index 
$\Gamma$ ($N(E)\propto E^{-\Gamma}$), the star name, the angular distance 
from the star to the $\gamma$-ray source best position, the distance to the 
star, the terminal wind velocity, the mass loss rate, the expected intrinsic 
$\gamma$-ray luminosity assuming the star's distance (the minimum one when 
there are more than one star in the field) and isotropic emission 
with average index $\Gamma=2$, and, in the last column, any other positional 
coincidence revealed in our study. 

From Table 2, it can be seen that most of the possible associations claimed by 
Raul \& Mitra (1997) are no longer viable ones. Just WR stars 37-39, 138, and
142 of their list stay after our analysis.    

In order to compare Raul and Mitra's results with our's, it is worth
remembering that when testing against positional coincidences they assumed
an angular uncertainty of 1$^o$ for all EGRET sources.
If we would make such an assumption, we would have found 20 positional
coincidences in the 3EG catalog (i.e. 24.7 \% of the unidentified
low-latitude sources). However, due to the new reduced EGRET
errors, just 7 \% of these sources are now positionally consistent with WR
stars, with a priori probability of $\sim10^{-3}$ of being by chance. In
addition, there are 4 sources with Of stars within their error boxes. The
probability that these latter associations result just by chance is
$\sim10^{-2}$. 

Several mechanisms have been proposed to generate $\gamma$-rays in the
vicinity of massive stars with strong winds. A compact $\gamma$-ray source
could be the result of $\pi^0$-decays which occur as a consequence of
hadronic interactions between relativistic protons or nuclei, locally
accelerated by
shocks arising from line-driven instabilities in the star wind, and
thermal ions (White \& Chen 1992). The same embedded shocks can also
accelerate electrons that could provide an additional source of (inverse
Compton) $\gamma$-ray emission through the upscattering of stellar UV
photons (Chen \& Withe 1991a). Synchrotron losses of these energetic
electrons can produce observable nonthermal radio emission, as detected in
several massive stars (e.g. Abbott et al. 1984). 

A different region where the $\gamma$-rays might be generated is at the
interface between the supersonic wind flow and the interstellar medium.
There, the terminal shock can reaccelerate ions up to high energies and, if
sufficient concentration of ambient matter is available (e.g. small clouds
or swept-up material), nuclear $\gamma$-rays copious enough to be detected
could be produced (Cass\'e \& Paul 1980). V\"olk \& Forman (1982) have
argued that stellar energetic particles lose too much energy in the
expanding wind to be efficiently accelerated at the terminal shock, in
such a way that local injection (e.g. from a nearby star) is required.
However, White (1985) showed that the shocks embedded in the highly
unstable radiatively driven winds can be responsible for much higher
initial energies and partial reacceleration of the particles during the
adiabatic expansion, so isolated massive stars could be also efficient
$\gamma$-ray emitters if they present sufficiently strong winds.

The a posteriori analysis of our association results show that three stars
are of especial interest as possible counterparts of EGRET sources: WR
140, WR 142, and Cyg OB2 No.5. The first one is a binary system composed of
a WC 7 plus an O4-5 star. The region of stellar wind collision seems to be
particularly suitable for producing high energy emission. Eichler \& Usov
(1993) have studied the particle acceleration in this system concluding that 
it should be a strong $\gamma$-ray source. Based on observational data on WR
140, they predicted a $\gamma$-ray luminosity in the range
$5\times10^{32}-2.5\times10^{35}$ erg s$^{-1}$, in well agreement with the
measured EGRET flux from 3EG J2022+4317 and the distance to the system
(see Table 2).

The second promising star, WR 142, is one of the five WR stars which
present strong OVI lines without being associated with planetary nebulae.
The large Doppler broadening of all spectral lines reveals the existence of
a very high wind velocity of $\sim5200$ km s$^{-1}$, which doubles what is
usually observed in WR stars (Polcaro et al. 1991). The identification of
WR 142 with the COS-B source 2CG 075+00 was proposed in Polcaro et al.'s 
(1991) paper, where they considered the $\gamma$-ray production in the strong stellar
wind. In the 3EG catalog the star position is consistent with the source
3EG J2021+3716. If the star is responsible for the observed $\gamma$-ray
flux, its intrinsic luminosity would be $\sim3\times10^{34}$ erg s$^{-1}$,
which is of the order of what is expected from White \& Chen's (1992)
hadronic model for isolated stars.

Finally, the binary system Cyg OB2 No.5 seems to be another interesting
candidate for producing $\gamma$-rays. Usually considered as a contact 
binary formed by two O7 I stars (e.g. Torres-Dodgen et al. 1991), recent 
observations suggest that the secondary star in this system would be of 
spectral type B0 V--B2 V (Contreras et al. 1997). Variable radio emission 
was detected by several authors (e.g. Persi et al. 1990), with 
timescales of $\sim7$ years. A weak radio component of nonthermal
nature has been observed with the VLA at a separation of $\sim0.8''$ 
from the main radio source, which is thermal and coincident with the 
primary optical component (Contreras et al. 1997). The radio 
variability in Cyg OB2 No.5 has been interpreted in terms of a colliding
wind model by Contreras et al. (1997), who suggested that the weaker radio
component is not a star but a bow shock produced by the wind collision. In
this shock, electrons can be locally accelerated up to relativistic energies,
yielding the synchrotron radiation that constitutes the secondary nonthermal
source. Additionally, $\gamma$-rays are generated through inverse Compton 
losses in the UV radiation field of the secondary (Eichler \& Usov 1993). 
The same Fermi mechanism that accelerates the electrons should also 
operate on protons, providing a source of energetic ions that could 
contribute with higher energy $\gamma$-ray emission, as in the case involving 
WR stars. For strong shocks, the test particle theory predicts that the
relativistic protons will have a differential energy spectrum 
given by $N(E)\propto E^2$. The $\pi^0$-decay $\gamma$-rays resulting from 
$p-p$ collisions should conserve the shape of the original proton spectrum, 
in such 
a way that at energies above 100 MeV the photon spectral index would be
$\Gamma\sim2$, as observed by EGRET.

\section{Supernova remnants}

Possible correlation between SNRs and unidentified EGRET sources,
on the basis of two dimensional positional coincidence, has been
proposed since the release of the first EGRET (1EG) catalog.
Sturner \& Dermer (1995) suggested that some of the unidentified
sources lying at galactic latitudes $|b|<10^o$ might be associated
with SNRs: of 37 detections, 13 overlapped SNR positions in the 1EG catalog. 
However, their own analysis
showed that the statistical significance was not too high as to 
provide a strong confidence. Chance association was just 1.8$\sigma$ away
from the obtained result. Using the 2EG catalog, Sturner et al. (1996) 
repeated the analysis, and showed that 95\%
confidence contours of 7 unidentified EGRET sources overlapped 
SNRs, some of them appearing to be in interaction with molecular clouds. 
Similar results were
independently reported by Esposito et al. (1996), although neither of them
assessed the chance probablility of these 2EG-catalog findings.
Considering the 1EG catalog, 35\% of the unidentified sources were positional
related to SNRs. This drop to 21.8\% in the 2EG catalog, and is currently
about 27\%. One important point to take into account when evaluating these 
differences is not only to consider the evolution of
the EGRET catalog but also that of the supernova remnant Green's catalog.
At the time of the first studies by Sturner \& Dermer (1995), the supernova 
catalog contained 182 SNRs. This grew up to 194 in 1996, and currently it
lists 220 remnants.

In Table 4 we show the 3EG sources that are positionally consistent with
SNRs listed in the latest version of Green's catalog. From left to right we
provide the $\gamma$-ray source name, the measured flux, the photon
spectral index $\Gamma$, the SNR identification, the angular distance
between the best $\gamma$-ray source position and the center of the remnant, the
size of the remnant in arcminutes, the SNR type (S for shell, F for
filled-centre, and C for composite), and other positional coincidences found in our
study. The table contains 22 possible associations with an a priori
probability of being purely by chance completely negligible ($\leq
10^{-8}$). It is important to remind that this list is formed entirely of
positional coincidences with currently catalogued SNRs. However, the
diffuse galactic disk nonthermal emission, originated in the interaction
of the leptonic component of cosmic rays with the galactic magnetic field,
is veiling many remnants of low surface brigthness. Recent observational
studies using filtering techniques in the analysis of radio data have
revealed many new SNR candidates that are not included in Green's catalog
(e.g. Duncan et al. 1995, Combi \& Romero 1998, Combi et al. 1998b, 1999). 
If these candidates
were included in our analysis a larger number of associations would have
resulted. 

The intrinsic $\gamma$-ray luminosity of SNRs, stemmed from interactions
between cosmic rays reaccelerated at the supernova shock front and
swept-up material, is expected to be rather low (Drury et al. 1994). However, if a cloud is
near the particle acceleration site, the enhanced nuclear cosmic rays from
the shock can ``illuminate'' the cloud through $\pi^0$-decays yielding a
compact $\gamma$-ray source (Aharonian et al. 1994). Such scenario has
been recently study by Combi et al (1998a) in relation with the source
3EG J1659-6251 (previously 2EGS J1703-6302).

\section{OB associations}

In Table 5 we list the unidentified 3EG sources that are positionally
coincident with the OB associations in the catalog by Mel'nik \& Efremov
(1995). Our results can be compared with the similar work by Kaaret \& Cottam
(1996).
Using the 2EG catalog, they have already 
found a statistically significant 
correlation: 9 of the unidentified 2EG sources 
have position contours overlapping
an OB association and other 7 lie within $1^o$ angular distance.
These results are totally compatible with our's.
Here, we find 26 superpositions out of 81 unidentified 3EG sources (32\%), 
5$\sigma$ away from what is expected from
pure chance association. The mean angular separation between the centroid
of the OB association and the EGRET source is 1.5$^o$, although most sources
are at angular distances of less than $1^o$.

The differences between both methods of analysis are worth commenting. 
In particular,
we decided, for completitude, to keep the nearby association Sco 2A despite
its proximity. Any pulsar traced by it must have a negligible proper
velocity in order to be consistent with its angular size, but its
existence cannot be ruled out only on a priori grounds. 
To calculate the chance superposition probability
Kaaret and Cottam studied EGRET sources just within [-5$^o$, 5$^o$] in 
galactic latitude
(only 25 sources of the total 129 unidentified ones present in the
2EG catalog),
and generated sample locations using two Gaussian distributions, in longitude
and latitude, 
with central value and deviation provided by the actual positions
of the unidentified sources. They also used a galactic model
to map the gas distribution.
This procedure yields almost the same results than the method we follow 
(chance association probability around $10^{-5}$).
Interestingly, despite
all EGRET sources changed their positions
from the 2EG to the 3EG catalog and a significant number of new
detections has
been added, the percentage and the confidence level of the positional
coincidences remains almost the same in both studies.

All known $\gamma$-ray pulsars are young objects ($\leq 10^{6}$ yr) with
spectral indices smaller than 2.15 (Crab's) and a trend for spectral
hardening with characteristic age (Fierro et al. 1993). From Table 5, 
if we consider just
sources coincident {\em only} with OB associations and exclude the three
sources with very steep indices (3EG J 1308-6112, 3EG J 1718-3313, and 3EG
J 1823-1314), we get $<\Gamma>=2.07$ and $1\sigma=0.12$ for the 8
remaining EGRET sources. These are the most promising candidates for
pulsar associations. We have marked them with a star symbol in Table 5.

\section{Further comments}

\begin{figure}
\resizebox{\hsize}{!}{\includegraphics{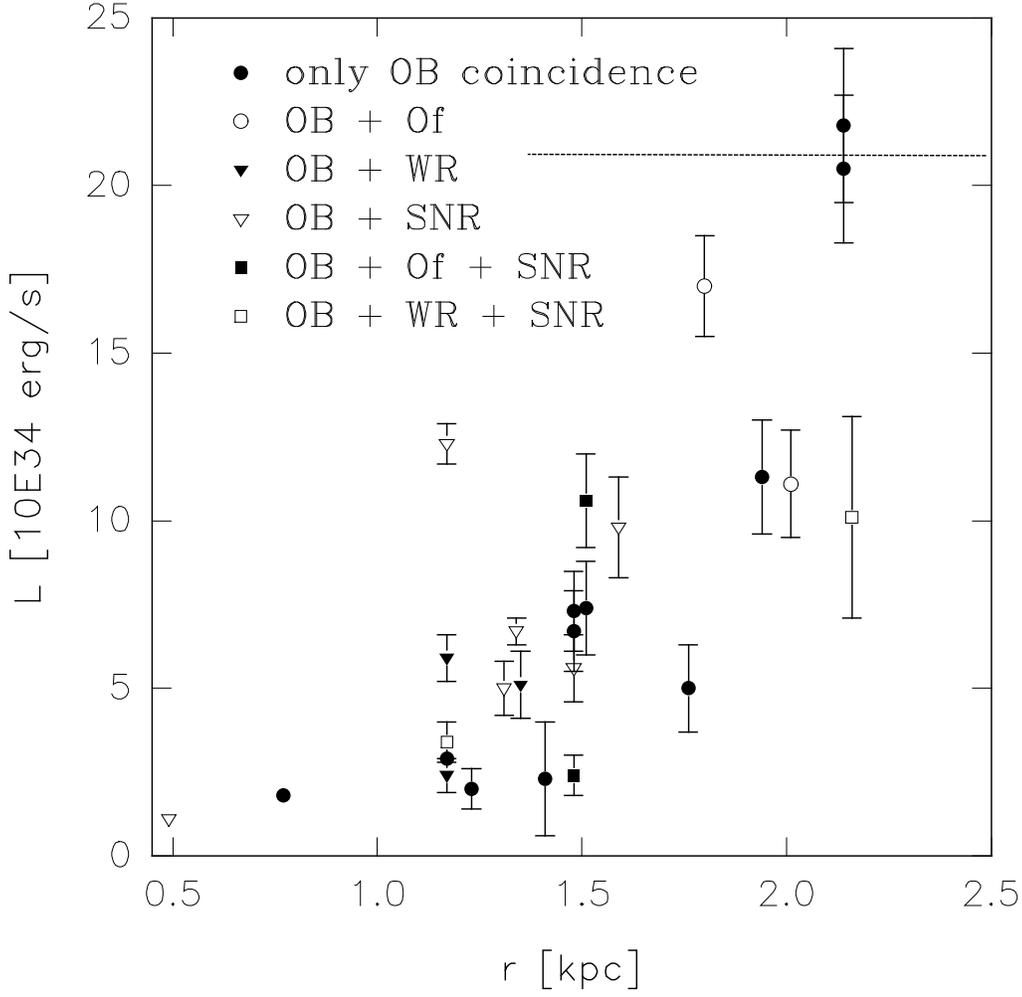}}
\caption{Luminosity versus distance for unidentified 3EG sources coincident 
{\em only} with OB associations (black filled dots) and with OB associations 
{\em and} other objects (remaining symbols). The horizontal line marks 
the luminosity of Vela pulsars. 
Isotropic emission has been considered in all cases.}
\label{fig.2}
\end{figure}

\begin{figure}
\resizebox{\hsize}{!}{\includegraphics{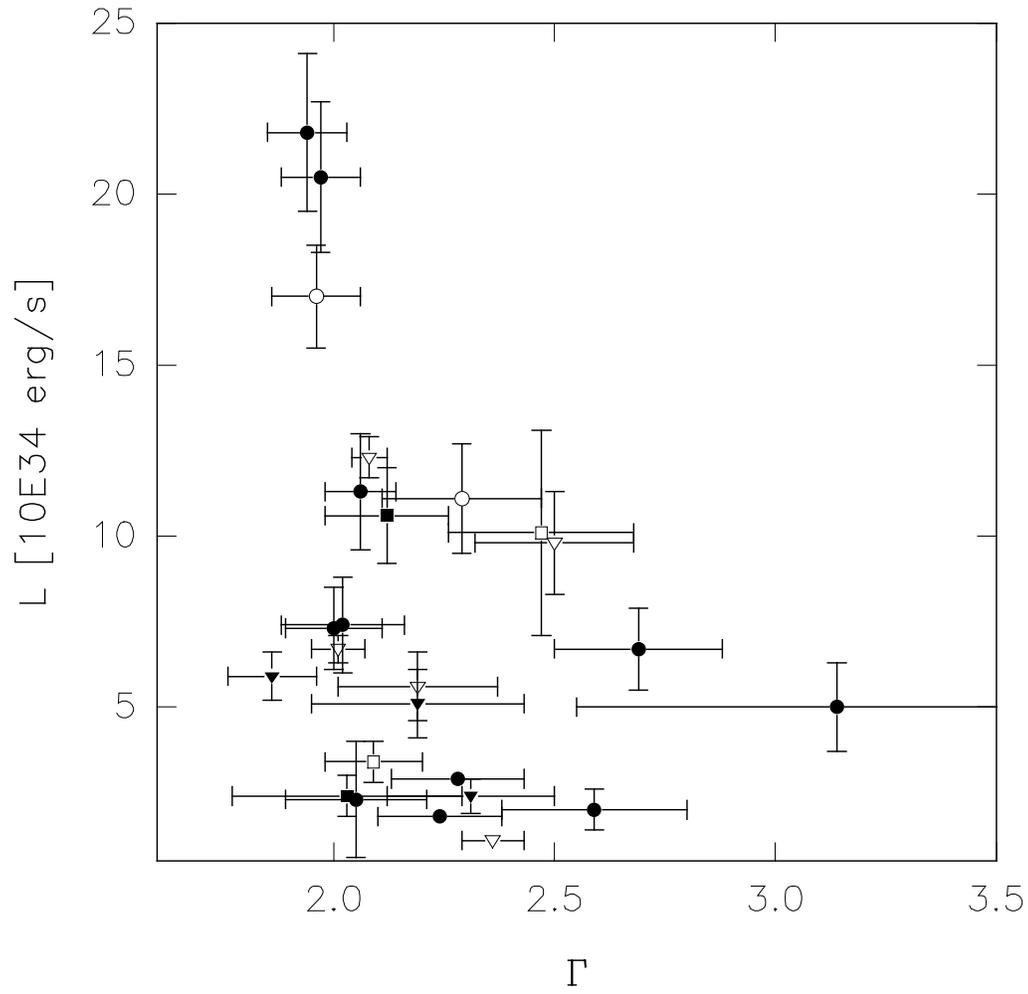}}
\caption{Luminosity versus photon spectral index for unidentified 3EG sources
coincident with OB associations. Isotropic emission is assumed in
all cases. Symbols are as in Fig.2.}
\label{fig.3}
\end{figure}

In Figure ~\ref{fig.2} we show a plot of the $\gamma$-ray luminosity (assuming isotropic 
emission) of the unidentified sources coincident with OB associations against 
the estimated distance to the associations. By using different symbols we 
indicate whether there are additional positionally coincident objects for 
each $\gamma$-ray source. The solid horizontal line represents the luminosity
of Vela pulsar. A similar plot of luminosity versus photon spectral index
$\Gamma$ is shown in Figure ~\ref{fig.3}. The first plot shows that the luminosity 
distribution of this subset of 3EG sources is consistent with the observed
distribution for $\gamma$-ray pulsars when emission into $4\pi$ sr is 
assumed (see Kareet \& Cottam 1996). Figure ~\ref{fig.3} shows, however, that not 
all sources superimposed to OB associations present the spectral signature 
expected from pulsars: they should concentrate in the left-upper corner of 
the frame. There, two sources clearly differentiate from the rest:
3EG J1027-5817 and 3EG J1048-5840. They have luminosities similar to Vela's
and hard spectra with $\Gamma<2$, which make them good candidates for 
$\gamma$-ray pulsars.

The identification of 3EG J1048-5840 (formerly 2EG J1049-5847) with a pulsar
(PSR B1046-58) was already proposed by Zhang \& Cheng (1998), who showed
that its $\gamma$-ray spectrum is consistent with the predictions of outer
gap models. In addition, these authors also suggested that 3EG J1823-1314 
(2EG J1825-1307) could be the pulsar PSR B1823-13. This latter identification must 
be now rejected in the light of the new determination of the spectral 
index of the $\gamma$-ray source in the 3EG catalog, $\Gamma=2.69\pm0.19$,
which is too steep for a pulsar. 

Regarding 3EG J1027-5817, no known radio
pulsar is found within its 95 \% confidence contour. It could be a 
Geminga-like object or the effect of the combined emission of a pulsar and
a weak SNR in Car 1A-B (the 3EG catalog notes that it is a possible case of 
multiple or extended source).

Some of the low luminosity sources in Fig. ~\ref{fig.3} might be yet undetected SNRs, 
whereas the sources with the steepest indices could be background AGNs. 
A simple extrapolation of the high latitude population of $\gamma$-ray 
blazars shows that about 10 of this sources should be detected throughout
the Galaxy within $|b|<10^o$ (Yadigaroglu \& Romani 1997). Most of them,
however, should belong to the group of 43 3EG sources for which we have 
found not positional coincidences with any known galactic object. This 
set of sources has an average value of galactic latitude $<|b|>=5.8\pm3.3$,
which suggests a significant extragalactic contribution.

Finally, we want to mention two interesting additional possibilities to
explain some 3EG sources: isolated Kerr-Newman black holes (Punsly 1998a,b)
and isolated standard black holes accreting from the diffuse interstellar
medium (Dermer 1997). In Punsly's model, a bipolar magnetically dominated 
MHD wind is driven by a charged black hole located in a low density region
(otherwise it would discharge rapidly). The wind forms two leptonic jets
which propagate along the rotation axis in opposite directions, as it 
occurs in AGNs. Self-Compton losses provide $\gamma$-ray luminosity in the
range $10^{32}-10^{33}$ erg s$^{-1}$ for a 7-$M_{\odot}$ black hole with a 
polar magnetic field of $\sim10^{10}$ G. If such an object is relatively
close ($\sim300$ pc), it could appear as a typical unidentified EGRET 
source with $\Gamma\sim2.5$.

In the case of isolated black holes accreting from a diffuse medium, a 
hole with mass of 10 $M_{\odot}$ and a velocity of 10 km s$^{-1}$ can 
produce a $\gamma$-luminosity $\sim7\times10^{33}$ erg s$^{-1}$ 
in a medium with 
density of 0.1 cm$^{-3}$ (Dermer 1997). Changes in the particle density can 
result in $\gamma$-ray flux variability, as observed in several unidentified 
sources. None of these $\gamma$-sources based on black holes can be ruled out at 
present, and their observational signatures at other wavelengths seem to
be worth of careful search.

\section{Conclusions}

We have studied the level of two-dimensional positional coincidences between 
unidentified EGRET sources at low galactic latitudes in the 3EG catalog 
and different populations of galactic objects, finding out that there is
overwhelming statistical evidence for the association of $\gamma$-ray sources
with SNRs and OB star forming regions (these latter considered as pulsar tracers). 
Additionally, there is marginally significant evidence for the association with 
early-type stars endowed with very strong winds, like Wolf-Rayet stars and Of stars.
A posteriori analyses of the star candidates show that there are at least 
three systems (WR 140, WR 142, and Cyg OB2 No. 5) which are likely 
$\gamma$-ray sources. Several sources positionally coincident with
OB associations are probably pulsars, like 3EG J1048-5840 and similar sources with
hard spectra. Besides, there are 43 3EG sources for which we have
not found any positional coincidence with known objects. This set of sources 
could include undetected low-brightness SNRs in interaction with dense 
and compact clouds, some Geminga-like pulsars, and, perhaps, a new kind of
galactic $\gamma$-ray sources, like Kerr-Newman black holes or isolated 
black holes accreting from the interstellar medium.

The main conclusion to be drawn is that there seems to exist more than a single
population of galactic $\gamma$-ray sources. Pulsars constitute a well 
established 
class of sources, and there is no doubt that under certain conditions some 
SNRs are also responsible for significant $\gamma$-ray emission in the EGRET
scope. Both isolated and binary early-type stars are likely to present high-energy
radiation strong enough to be detected by EGRET in some special cases. We propose
that, in addition to the well-known WR stars 140 and 142, the Cyg OB2 No. 5 binary 
system could be a strong $\gamma$-ray source, the first one to be detected 
involving no WR stars. The large number of unidentified EGRET sources free of any 
positional 
coincidence with luminous objects also encourage further studies to find whether
there exist a population of exotic objects yet undetected at lower wavelengths.

\begin{acknowledgements}

This work has been partially supported by the Argentine agencies CONICET
and ANPCT.

\end{acknowledgements}

\newpage

\begin{table*}
\caption[]{Statistical results.}
\begin{flushleft}
\begin{tabular}{l c c c c c}
\noalign{\smallskip}
\hline
\noalign{\smallskip}
Object & Actual & Simulated & Probability & Simulated & Probability \cr
type & coincidence &1$^o$-bin & 1$^o$-bin & 2$^o$-bin & 2$^o$-bin\cr
\noalign{\smallskip}
\hline
\noalign{\smallskip}
WR & 6 & $2.3 \pm 1.4$  & $8.3 \times 10^{-3}$ & $2.1 \pm 1.4$ & $5.8\times 
10^{-3}$\cr
Of & 4 & $1.2 \pm 1.1$  & $1.5\times 10^{-2}$ & $1.1 \pm 1.0$ & $5.9\times 
10^{-3}$\cr 
Assoc.OB & 26 & $12.7 \pm 3.1$  & $1.2 \times 10^{-5}$ & $12.5 \pm 3.1$ & $9.8 \times 10^{-6}$\cr      
SNR & 22 & 7.8 $\pm$ 2.5  & $1.6 \times 10^{-8}$ & 7.0 $\pm$ 2.4 
& $5.4 \times 10^{-10}$ \cr
\noalign{\smallskip}
\hline
\end{tabular}
\end{flushleft}
\end{table*}

\begin{table*}
\caption[]{Positional coincidences with WR stars.}
\begin{flushleft}
\begin{tabular}{l c c l c c c c c l}
\hline
\noalign{\smallskip}
$\gamma$-Source & $F_{\gamma} \times 10^{8}$  & $\Gamma$ & Star & 
$\Delta\theta$ & $r$ & $v_{\infty}$ & $\log$\.M & $L \times10^{-34}$& Other\cr 
(3EG J)& (ph cm$^{-2}$ s$^{-1}$) & & & (deg) & (kpc) & (km s$^{-1}$) 
&(M$_{\odot}$ yr$^{-1}$) & (erg s$^{-1}$)& coincidences \\
\noalign{\smallskip}
\hline
\noalign{\smallskip}
0747-3412 & 16.3$\pm$5.0&2.22$\pm$0.18 & WR 9 (B) & 0.37 & 2.35 &
2200$^{\rm a}$ & -4.2$^{\rm f}$ & 6.5$\pm$ 2.0 &\\
1102-6103 & 32.5$\pm$6.2&2.47$\pm$0.21 & WR 34 & 0.48 & 9.50 &1200$^{\rm b}$ 
& -4.5$^{\rm b}$&&OB/SNR\\
 &   &   & WR 35     &  0.30 & 9.58 & 1100$^{\rm b}$ & -4.3$^{\rm b}$ &&\\
 &   &   & WR 37     &  0.45 & 2.49 & 2150$^{\rm b}$ & $<$-4.1$^{\rm f}$&&\\
 &   &   & WR 38     &  0.45 & 1.97 & 2400$^{\rm a}$ & $<$-4.2$^{\rm f}$ &&\\
 &   &   & WR 39     &  0.51 & 1.61 & 3600$^{\rm a}$ & $<$-4.0$^{\rm f}$
& 6.1$\pm$1.1&\\
1655-4554 & 38.5$\pm$7.7&2.19$\pm$0.24 & WR 80 & 0.59 & 4.40 & 2000$^{\rm c}$ 
& -4.1$^{\rm c}$&54.0$\pm$10.8 & OB\\
2016+3657 & 34.7$\pm$5.7&2.09$\pm$0.11 & WR 137 (B)& 0.44 & 1.82 & 
1900$^{\rm d}$ & -4.5$^{\rm d}$& 8.3 $\pm$1.4&OB/SNR\\
 &   &   & WR 138    &  0.50 & 1.82 & 1500$^{\rm b}$ & -4.7$^{\rm d}$&\\
2021+3716 & 59.1$\pm$6.2&1.86$\pm$0.10 & WR 142& 0.15 & 0.95 & 5200$^{\rm e}$ 
& $<$-4.7$^{\rm g}$& 3.9$\pm$0.4&OB\\
2022+4317 & 24.7$\pm$5.2&2.31$\pm$0.10 & WR 140 (B)& 0.64 & 1.34 & 2900$^{\rm d}$ 
& -4.1$^{\rm d}$& 3.2 $\pm$0.7&OB\\
\noalign{\smallskip}
\hline
\multicolumn{10}{l}
{a: Koesterke \& Hamann (1995); b: Hamann \& Koesterke (1998); c: Zubko (1995); d: Prinja
et al. (1990); e: Polcaro et al. (1991);}\cr
\multicolumn{10}{l}
{f: Leitherer et al. (1997); g: Abbott et al. (1986); (B) stands for ``binary system''.}\cr
\end{tabular}
\end{flushleft}
\end{table*}

\begin{table*}
\caption[]{Positional coincidences with Of stars.}
\begin{flushleft}
\begin{tabular}{l c c l c c c c c l}
\noalign{\smallskip}
\hline
\noalign{\smallskip}
$\gamma$-Source & $F_{\gamma} \times 10^{8}$  & $\Gamma$ & Star & 
$\Delta\theta$ & $r$ & $v_{\infty}$ & $\log$\.M & $L \times10^{-34}$& Other\cr 
(3EG J)& (ph cm$^{-2}$ s$^{-1}$) & & & (deg) & (kpc) & (km s$^{-1}$) 
&(M$_{\odot}$ yr$^{-1}$) & (erg s$^{-1}$)& coincidences \\
\noalign{\smallskip}
\hline
\noalign{\smallskip}
0229+6151 & 37.9$\pm$6.2 & 2.29$\pm$0.18 & HD 15629  &  0.30 &1.9$^{\rm a}$ 
&2900$^{\rm c}$ & -5.8$^{\rm c}$ & 10 $\pm$ 1& OB\cr
0634+0521 & 15.0$\pm$3.5 & 2.03$\pm$0.26& HD 46150  &  0.64 & 1.3$^{\rm a}$  
&2900$^{\rm c}$  & $<$-5.9$^{\rm c}$ & 1.8 $\pm$ 0.5 & OB/SNR\cr
      &   &   & HD 46223 &  0.63 & 1.6$^{\rm a}$ & 2800$^{\rm c}$ &
-5.8$^{\rm c}$&& \cr 
1410-6147 & 64.2$\pm$8.8& 2.12$\pm$0.14  & HD 124314 &  0.25 & 1.0$^{\rm a}$ 
& 2400$^{\rm d}$ & -4.7$^{\rm g}$ & 4.6 $\pm$ 0.7 & OB/SNR\cr
2033+4118  &  73.0$\pm$6.7& 1.96$\pm$0.10 & Cyg OB2 5 (B)&  0.27 & 
1.8$^{\rm b}$ & 1500$^{\rm e}$ & -4.43$^{\rm e}$ & 17 $\pm$ 1.5 & OB\cr
      &   &   & Cyg OB2 11&  0.24 &1.8$^{\rm b}$   & 2500$^{\rm f}$ & 
-5.2$^{\rm h}$ &&\cr
\noalign{\smallskip}
\hline
\noalign{\smallskip}
\multicolumn{10}{l}
{a: computed using $M_v$ from Vacca et al. (1996); b: Humphreys (1978); 
c: Leitherer et al. (1997); d: Prinja et al. (1990);}\cr
\multicolumn{10}{l}
{e: Conti \& Howarth (1999); f: Bieging et al. (1989); g: Benaglia et al. (1998); 
h: computed as in Lamers \& Leitherer (1993);}\cr
\multicolumn{10}{l}
{(B) stands for ``binary system''.}\cr
\end{tabular}
\end{flushleft}
\end{table*}

\clearpage

\begin{table*}
\caption[]{Positional coincidences with supernova remnants.}
\begin{flushleft}
\begin{tabular}{l c c l c c c l}
\noalign{\smallskip}
\hline
\noalign{\smallskip}
$\gamma$-source  & $F_{\gamma} $ & $\Gamma$ & ${\rm SNR}$  & $\Delta\theta$ 
&  Size & Type & Other \\
(3EG J)  &($10^{-8}$ ph cm$^{-2}$ s$^{-1}$)& & & (deg) & (arcmin) & &
coincidences\\      
\noalign{\smallskip}
\hline
\noalign{\smallskip}
0542+2610 & 14.7$\pm$3.2 & 2.67$\pm$0.22 & G180.0-1.7         & 2.04& 180    
&  S &  \\
0617+2238 & 51.4$\pm$3.5 & 2.01$\pm$0.06 & G189.1+3.0         & 0.11& 45    
&  S  & OB \\
0631+0642 & 14.3$\pm$3.4 & 2.06$\pm$0.15 & G205.5+0.5         & 1.97& 220    
&  S  & \\
0634+0521 & 15.0$\pm$3.5 & 2.03$\pm$0.26 & G205.5+0.5         & 2.03& 220    
&  S &  Of/OB\\
0824-4610 & 63.9$\pm$7.4 & 2.36$\pm$0.07 & G263.9-3.3         & 1.71& 255    
&   C  & OB \\
0827-4247 & 42.6$\pm$7.4 & 2.10$\pm$0.12 & G260.4-3.4         & 1.04& 60 
$\times 50$    &   S & \\
0841-4356 & 47.5$\pm$9.3 & 2.15$\pm$0.09 & G263.9-3.3         & 2.28& 255    
&  C  & \\
1013-5915 & 33.4$\pm$6.0 & 2.32$\pm$0.13 & G284.3-1.8         & 0.65& 24    
&   S & \\
1102-6103 & 32.5$\pm$6.2 & 2.47$\pm$0.21 & G290.1-0.8         & 0.12& 19
$\times 14$ & S   & WR/OB    \\
1410-6147 & 64.2$\pm$8.8 & 2.12$\pm$0.14 & G312.4-0.4         & 0.23&     
&  S&  OB/Of  \\
1639-4702 & 53.2$\pm$8.7 & 2.5$\pm$0.18  & G337.8-0.1         & 0.07& 9
$\times 6$  & S   & OB     \\
          &               &              & G338.1+0.4         & 0.65& 15    
&  S  & \\
          &               &              & G338.3+0.0         & 0.57& 8    
&  S  & \\
1714-3857 & 43.6$\pm$6.5 & 2.30$\pm$0.20 & G348.5+0.0         & 0.47& 10    
&  S  & \\
          &               &              & G348.5+0.1         & 0.50&  15   
&  S &  \\
1734-3232 & 40.3$\pm$6.7 &     $-$       & G355.6+0.0         & 0.16& 6
$\times 8$    & S  & OB  \\
1744-3011 & 63.9$\pm$7.1 & 2.17$\pm$0.08 & G359.0-0.9         & 0.41& 23    
&   S & \\
          &              &               & G359.1-0.5         & 0.25&  24   
&  S  &  \\
1746-2851 &119.9$\pm$7.4 & 1.70$\pm$0.07 & G0.0+0.0           & 0.12& 3.5
$\times 2.5$    & S &   \\
          &              &               & G0.3+0.0           & 0.19& 15.8    
&  S  & \\
1800-2338 & 61.3$\pm$6.7 & 2.10$\pm$0.10 & G6.4-0.1          & 0.17&  42   
& C   & \\
1824-1514 & 35.2$\pm$6.5 & 2.19$\pm$0.18 & G16.8-1.10         & 0.43&  30
$\times 24$   &  & OB  \\
1837-0423 & 19.1         & 2.71$\pm$0.44 & G27.8+0.6        & 0.58& 50
$\times 30$    & F   & \\
1856+0114 & 67.5$\pm$8.6 & 1.93$\pm$0.10 & G34.7-0.4         & 0.17& 35
$\times 27$    &   S & \\
1903+0550 & 62.1$\pm$8.9 & 2.38$\pm$0.17 & G39.2-0.3         & 0.41& 8
$\times 6$    &  S  & \\
2016+3657  & 34.7$\pm$5.7 & 2.09$\pm$0.11 & G74.9+1.2          & 0.26& 8
$\times 6$    &  F & WR/OB \\
2020+4017  &123.7$\pm$6.7 & 2.08$\pm$0.04 & G78.2+2.1         & 0.15&  60   
&  S  & OB\\
\noalign{\smallskip}
\hline
\noalign{\smallskip}
\end{tabular}
\end{flushleft}
\end{table*}

\begin{table*}
\begin{flushleft}
\caption[]{Positional coincidences with OB associations.}
\begin{tabular}{l c c l c c c c l}
\noalign{\smallskip}
\hline
\noalign{\smallskip}
$\gamma$-source & $F_{\gamma} $ & $\Gamma$ & OB Assoc. & $\Delta\theta$ 
& $r$  & Size & $L $ & Other\\
(3EG J)&($10^{-8}$ ph cm$^{-2}$ s$^{-1}$)& & & (deg) & (kpc) & (deg)
 &  ($10^{34}$ erg s$^{-1}$)& coincidences\\      
\hline
0229+6151 & 37.9$\pm$6.2 & 2.29$\pm$0.18 &Cas 6     & 1.44& 2.01& 0.95 & 
11.1 $\pm$ 1.6 & Of\\
0617+2238 & 51.4$\pm$3.5 & 2.01$\pm$0.06 &Gem 1     & 2.07& 1.34& 1.94& 6.7 
$\pm$0.4 & SNR\\
0634+0521 & 15.0$\pm$3.5 & 2.03$\pm$0.26 &Mon 2A    & 1.27& 1.63& 0.84& &
Of/SNR\\
    &   &    &Mon 1B    & 1.27& 1.48& 0.60 & 2.4 $\pm$ 0.6&\\
0824-4610 & 63.9$\pm$7.4 & 2.36$\pm$0.07 &VELA 2    & 4.75& 0.49& 4.14& 
1.1 $\pm$ 0.1 & SNR\\
0848-4429$^\star$ & 20.1$\pm$7.7 & 2.05$\pm$0.16 &VELA 1B   & 1.62& 1.41& 1.00& 
2.3 $\pm$ 1.7 &\\
1027-5817$^\star$ & 65.9$\pm$7.0 & 1.94$\pm$0.09 &Car 1A    & 1.17& 2.40& 0.80&&\\
        &  &  &Car 1B    & 1.98& 2.14& 1.61& 21.8 $\pm$2.3&\\
1048-5840$^\star$ & 61.8$\pm$6.7 & 1.97$\pm$0.09 &Car 1B    & 1.78& 2.14& 1.61&
20.5$\pm$ 2.2 &\\
       &     &    &Car 1E    & 1.72& 2.64& 1.55& &\\
       &     &    &Car 1F    & 0.72& 2.76& 0.55& &\\
1102-6103 & 32.5$\pm$6.2 & 2.47$\pm$0.21 &Car 1-2   & 1.17& 2.62& 0.56& & 
WR/SNR\\
       &     &    &Car 2     & 3.15& 2.16& 2.54& 10.1 $\pm$ 3.0&\\
1308-6112 & 22.0$\pm$6.1 & 3.14$\pm$0.59 &Cen 1B    & 1.15& 1.76& 0.44& 
5.0 $\pm$1.3 &\\
       &     &    &Cen 1D    & 1.46& 1.87& 0.75&&\\
       &     &    &Sco 2A    & 8.89& 0.16& 8.18& 0.04 $\pm$0.01&\\
1410-6147 & 64.2$\pm$8.8 & 2.12$\pm$0.14 &CLUST 3   & 1.12& 1.51& 0.76& 
10.6 $\pm$1.4 & Of/SNR\\
1420-6038$^\star$ & 44.7$\pm$8.6 & 2.02$\pm$0.14 &CLUST 3   & 1.08& 1.51& 0.76& 7.4 
$\pm$ 1.4&\\
1639-4702 & 53.2$\pm$8.7 & 2.5$\pm$0.18  &Ara 1A A  & 1.95& 1.59& 1.39& 
9.8 $\pm$ 1.5&SNR\\
       &     &    &NGC 6204  & 2.11& 1.94& 1.55&&\\
1655-4554 & 38.5$\pm$7.7 & 2.19$\pm$0.24 &Ara 1A B  & 1.13& 1.35& 0.47& 
5.1 $\pm$ 1.0 & WR\\
1718-3313 & 18.7$\pm$5.1 & 2.59$\pm$0.21 &Sco 4     & 1.84& 1.23& 1.30  & 
2.0 $\pm$0.6&\\
1734-3232 & 40.3$\pm$6.7 &     $-$       &Tr 27     & 1.63& 1.31& 1.14& 
5.0 $\pm$0.8 & SNR\\
1809-2328$^\star$ & 41.7$\pm$5.6 & 2.06$\pm$0.08 &Sgr 1B    & 0.47& 1.94& 0.31 & 
11.3 $\pm$ 1.7&\\
1823-1314 & 42.0$\pm$7.4 & 2.69$\pm$0.19 &Sct 3     & 1.45& 1.48& 1.16& 
6.7 $\pm$ 1.2 &\\
1824-1514 & 35.2$\pm$6.5 & 2.19$\pm$0.18 &Sct 3     & 1.68& 1.48& 1.16& 
5.6 $\pm$ 1.0 & SNR\\
1826-1302$^\star$ & 46.3$\pm$7.3 & 2.00$\pm$0.11 &Sct 3     & 1.62& 1.48& 1.16& 
7.3 $\pm$ 1.2 &\\
2016+3657 & 34.7$\pm$5.7 & 2.09$\pm$0.11 &Cyg 1,8,9 & 5.49& 1.17& 4.94& 
3.4 $\pm$0.6 & WR/SNR\\
2020+4017 &123.7$\pm$6.7 & 2.08$\pm$0.04 &Cyg 1,8,9 & 5.10& 1.17& 4.94 & 
12.3 $\pm$0.6 & SNR\\
2021+3716 & 59.1$\pm$6.2 & 1.86$\pm$0.10 &Cyg 1,8,9 & 5.24& 1.17& 4.94& 
5.9 $\pm$ 0.7 & WR\\
2022+4317 & 24.7$\pm$5.2 & 2.31$\pm$0.19 &Cyg 1,8,9 & 5.66& 1.17& 4.94& 
2.4 $\pm$0.5 & WR\\
2027+3429$^\star$ & 25.9$\pm$4.7 & 2.28$\pm$0.15 &Cyg 1,8,9 & 5.71& 1.17& 4.94& 
2.9 $\pm$0.1 &\\
2033+4118 & 73.0$\pm$6.7 & 1.96$\pm$0.10 &Cyg 1,8,9 & 5.22& 1.17& 4.94& 
7.2 $\pm$0.7 & Of\\
2227+6112$^\star$ & 41.3$\pm$6.1 & 2.24$\pm$0.14 &Cep 2 B   & 4.06& 0.77& 3.60& 
1.8 $\pm$0.2 &\\
\noalign{\smallskip}
\hline
\noalign{\smallskip}
\multicolumn{9}{l}
{$^\star$ Pulsar candidate?.}\cr
\end{tabular}
\end{flushleft}
\end{table*}

\newpage

\newpage

\newpage

\end{document}